\documentclass[preprint,12pt,onecolumn]{elsarticle}
\usepackage{graphicx,subfigure}
\usepackage{amsmath,amssymb}
\usepackage{float}
\usepackage[hidelinks]{hyperref}
\usepackage{bm}

\newcommand{\be}{\begin{eqnarray}}
\newcommand{\ee}{\end{eqnarray}}

\def\slashchar#1{\setbox0=\hbox{$#1$}           
   \dimen0=\wd0                                 
   \setbox1=\hbox{/} \dimen1=\wd1               
  \ifdim\dimen0>\dimen1                        
 \rlap{\hbox to \dimen0{\hfil/\hfil}}      
  #1                                        
 \else                                        
    \rlap{\hbox to \dimen1{\hfil$#1$\hfil}}   
    /                                         
 \fi}                                         %

\begin{document}

\title{The topological objects near the chiral crossover transition in QCD }

\author[1]{ Rasmus N. Larsen}
\address[1]{Physics Department, Brookhaven National Laboratory, Upton NY 11973}
\author[2]{Sayantan Sharma}
\address[2]{The Institute of Mathematical Sciences, Chennai 600113}
\author[3]{ Edward  Shuryak }
\address[3]{Physics Department, Stony Brook University, Stony Brook NY 11794 }

\begin{abstract}
We study the underlying topology of gauge fields in 2+1 flavor QCD with domain wall fermions on 
lattices of size $32^3\times 8$, at and immediately above the chiral crossover transition. 
Using valence overlap fermions with exact index theorem, we focus on its zero modes for different choices of periodicity phases along the temporal direction. Our studies show that the zero modes are due to fractionally charged topological objects, the instanton-dyons. We further provide qualitative study of the interactions between those and compare with the available semi-classical results, finding remarkably accurate agreement in all cases.  
\end{abstract}
\maketitle

\section{Introduction}
Chiral symmetry breaking and confinement are two of the most striking
non-perturbative phenomena that explains the phase diagram of strongly 
interacting matter described by Quantum Chromodynamics (QCD). While 
increasing evidences point to the fact that interacting topological 
objects in QCD -- monopoles, instantons, instanton-dyons are driving 
the corresponding phase transitions, quantitative understanding of 
their exact role still require a lot of  work.

Historically, discussion of chiral symmetry breaking started from Nambu-Jona-Lasinio (NJL)
model \cite{Nambu:1961tp}, in which hypothetical strong attraction between fermions have 
been introduced.  Based on analogy to theory of superconductivity, it explained formation 
of effective quark masses and massless pions. Two decades later the instanton liquid 
model (ILM) \cite{Shuryak:1981ff} identified the non-perturbative interaction with 
the instanton-induced 't Hooft Lagrangian. Unlike the 4-fermion interaction in the NJL model, 
it is not always an attractive thus explicitly violating the $U_A(1)$ symmetry. 
The ILM also proposed another view on the chiral symmetry breaking, related with
{\em collectivization} of the topological fermionic zero modes into the so called
zero-mode zone. Multiple numerical simulations of ensemble of interacting 
instantons were done, for a  review see~\cite{Schafer:1996wv}, which were able to 
reproduce point-to-point correlation functions corresponding to different mesons and 
baryons, known from phenomenology and lattice studies.

Relating these developments to confinement, it was shown in Ref. \cite{Kraan:1998sn,Lee:1998bb} 
that the nonzero holonomy (or the average Polyakov loop which is the confinement order 
parameter) can split $SU(N_c)$ instantons into $N_c$ constituents, known as 
instanton-dyons or instanton-monopoles. Henceforth we refer to them as dyons for simplicity. Ensembles of dyons via semi-classical methods were studied analytically~\cite{Liu:2015ufa,Liu:2015jsa,Liu:2016thw,Liu:2016yij} as well as numerically~\cite{Larsen:2015vaa,Larsen:2015tso,Larsen:2016fvs}. These works could reproduce both deconfinement and chiral phase transitions occurring at the same temperature in $N_c=2$ QCD, and also explain extra phase transitions in beyond QCD theories with modified quark periodicity phases by ``jumps" of the zero modes, from one type of dyon to another.

Two important questions need to be addressed at this stage. Firstly whether dyons 
exist in the QCD medium at finite temperature and secondly whether the topological properties of the QCD vacuum near the chiral crossover transition can be indeed explained in terms of a semi-classical theory of dyons. Non-perturbative nature of QCD adds to the enormous complexity of these problems with a quantitative or even a qualitative first-principles solution is extremely challenging. One of the early studies to address the first issue was done by Gattringer~\cite{Gattringer:2002wh} who used QCD Dirac operator with a reasonably good chiral property on the lattice and two different temporal periodicity conditions on $SU(3)$ pure 
gauge configurations as a tool to locate dyons. 
These results reported the existence of dyons in gauge theories without fermions. Further 
lattice studies have been performed by Mueller-Preussker, Ilgenfritz and collaborators~\cite{Bornyakov:2015xao,Bornyakov:2016ccq}, along similar lines. Clusters of topological fluctuations were identified using a local definition of topological charge, from the eigenvectors of the valence Dirac operator with generalized periodicity conditions. 
Observed correlation between the topological clusters and local eigenvalues of the 
Polyakov loop provided further evidence for the presence of dyons in QCD vacuum.  
However no non-perturbative studies, either analytic or in lattice field 
theory have been carried out on clarifying the possible existence of a semi-classical 
description of QCD vacuum near the deconfinement transition. Only recently, it has been established through lattice studies that at very high temperatures deep in the deconfined phase, the QCD vacuum can be described quite accurately in terms 
of a dilute gas of instantons~\cite{Petreczky:2016vrs,Frison:2016vuc,Borsanyi:2016ksw,Burger:2018fvb,Bonati:2018blm}.

This Letter is aimed at a more detailed and comprehensive understanding of the topological constituents of QCD with physical quark masses, near the chiral crossover transition temperature $T_c$. In particular, it is the first distinct attempt towards quantifying the density and interactions between different species of dyons as a function of temperature. We use the so called overlap Dirac operator~\cite{Narayanan:1994gw,Neuberger:1998my}, a particular realization of fermions on the lattice that has exact chiral symmetry, and therefore an exact index theorem~\cite{Hasenfratz:1998ri}, as detailed in the next section. This allows us to unambiguously identify the fermionic zero-modes and relate them to the underlying topological structures in QCD. We track their location and space-time profiles, by varying the temporal periodicity conditions for the overlap Dirac operator. In the subsequent sections, we provide convincing evidence that the topological objects that we identify in QCD near $T_c$ do indeed quite accurately correspond to expectations from the semi-classical theory of dyons.

\section{Methodology}
The configurations used in this work are $2+1$ flavor thermal QCD configurations generated by the HotQCD 
collaboration~\cite{Bhattacharya:2014ara} using M\"{o}bious domain wall discretization~\cite{Kaplan:1992bt,Brower:2012vk} 
for fermions and Iwasaki gauge action. 
The Euclidean space-time torus (lattice) has $N_s=32$ sites along the spatial directions and $N_\tau=8$ sites along the temporal direction.  The light and the strange quark masses are chosen to be physical corresponding to pion mass of $135$ MeV. The typical spatial size of the lattice is $\sim 4$ fm,  about four times the pion Compton wavelength. The temperature is defined by $T=1/(N_\tau a)$ where $a$ is the lattice spacing, $\sim 0.1$ fm. These same configurations 
have been previously used to calculate the chiral crossover transition temperature $T_c$ and investigate the $U_A(1)$ breaking~\cite{Bhattacharya:2014ara}. The pseudo-critical temperature was measured to be $T_c=155(9)$ MeV. The choice of M\"{o}bius domain wall fermions is crucial since it allows extremely accurate chiral symmetry on the lattice; its residual breaking is of the order of $\sim 2\times 10^{-3}$. 
We have chosen 13-14 statistically independent gauge configurations at each of the two temperatures $T/T_c \sim 1.0, 1.08$ respectively, where the expectation value of the Polyakov loop is intermediate between zero and unity. We have specifically selected configurations with topological charge $|Q_{top}|=1$, the cleanest set-up to study both zero and non-zero 
modes. Since the main aim of the paper is to identify the topological objects generating these modes, rather than calculating some average observables that require sampling of different $Q$ sectors, we think this choice is well justified.

Since the domain wall configurations have a small chiral symmetry breaking and does not satisfy an exact index theorem on the lattice,  we used zero modes of valence or probe overlap Dirac operator to detect the topological structures of these configurations. The overlap operator is defined as $D = 1-\gamma_5 sign(H_W)$ where the kernel of the sign function is $H_W = \gamma _5 (M - a D_W)$, $D_W$ being the massless Wilson-Dirac operator. $M$ is the domain wall height which is chosen to be in the interval $[0,2)$ to simulate one massless quark flavor on the lattice. Recall that the overlap operator satisfies the Ginsparg-Wilson relation~\cite{Ginsparg:1981bj}, $\gamma _5 D^{-1} + D^{-1} \gamma _5 = a \gamma _5$ defining the 'chiral invariance' on a finite lattice. Furthermore it has an exact index theorem, even at finite lattice spacings~\cite{Hasenfratz:1998ri} and hence its zero modes can be identified with the topological objects of the underlying gauge configurations. We have ensured that the overlap operator is implemented with good numerical precision. For all the configurations we have studied, the Ginsparg-Wilson relation and the sign function of the overlap operator was realized numerically with a precision of $10^{-9}$ or lower.

\subsection{Detecting the precise nature of the topological objects} 
This is done by varying temporal periodicity conditions for the probe fermion fields. 
We remind that the expectation value of the $SU(3)$ Polyakov loop is described by three eigenvalues at spatial infinity, known as the ``holonomy phases" $\mu_i$, which  divide the phase circle into $3$ different sectors of angular sizes $\nu_i=(\mu_{i+1}-\mu_i)/(2\pi)~,i=1,2,3$. The masses of the three types of dyons are proportional to $\nu_i$ 
and are called the $M_1, L$ and $M_2$ dyons corresponding to $\nu_{1,2,3}$ respectively.
For $SU(3)$ color group these phases in the confining phase are $\mu_1=0, \mu_2= 2\pi/3,\mu_3= 4\pi/3$ respectively, and therefore masses of all three types of dyons are the same. Since $\sum_i \nu_i=1$, the sum of masses of three types of dyons makes the instanton 
action. Still the Dirac operator should only have one zero-mode, per value of the topological charge $Q$. This means that only one of the dyons actually has the zero-mode: which one depends on the periodicity condition of the fermions on the Matsubara circle, $ \psi_f(\tau+\beta)=e^{i \phi} \psi_f(\tau)$.  Specifically, it is the dyon within the angular sector $\mu_i$ to $\mu_{i+i}$ to which the fermion phase $\phi$ belongs, that contains the zero-mode.  Now, if the topological constituents of the vacuum are a set of calorons i.e. finite temperature instantons, changing $\phi$ would not produce any changes in Dirac eigenstates. 
If on the other hand the vacuum consists of independent dyons, one would see ``jumps" as $\phi$ crosses $\mu_i$ and moves from one angular sector to the next. Note that while fermion phases $\phi$ are temperature independent, the holonomies $\mu_i$ move converging to zero at high temperature; therefore such jumps should lead to phase transition at a certain $T_c$.
  
In this work, the valence quark periodicity phases are chosen to be $\phi=-\pi/3,\pi/3,\pi$, the 
latter corresponding to the usual anti-periodic Matsubara periodicity condition. We then calculated the first 
$5$ eigenvalues and eigenvectors of the valence overlap Dirac operator on domain wall configurations using the 
Kalkreuter-Simma Ritz algorithm~\cite{Kalkreuter:1995mm}. The eigenvectors very determined with good numerical 
accuracy such that the error on the corresponding eigenvalues are $< 0.1\%$.
Our observables are two gauge invariant quantities, the 
wavefunction density $\rho(x)=\sum_{a=1}^{3}\sum_{i=1}^{4}\psi^\dagger_{a,i}(x)\psi_{a,i}(x)$ 
and the chiral density $\rho_5(x)=\sum_{a=1}^{3}\sum_{i,j=1}^{4}\psi ^\dagger_{a,i}(x)\gamma _{5(i,j)}\psi_{a,j}(x)$, 
defined in terms of the eigenvectors $\psi$ of the overlap operator. 
In subsequent sections, by studying space-time profiles of the zero modes and the ``jumps" of their positions 
for different periodicity conditions, we will identify the presence of dyons. The exact zero eigenmodes we study 
are well disentangled from the higher modes, making them completely insensitive to any ultra-violet fluctuations 
of the gauge fields. Furthermore, dyons are characterized by the holonomies i.e eigenvalues 
of the Polyakov loop so their space profiles would also have local lumps at the coordinates where we observe 
a peak in the fermion zero modes. However the gauge observables are much more noisier 
and require proper filtering of ultraviolet contamination, which is instead done more efficiently using the fermionic 
observables. A more detailed study on the robustness of our results with respect to different lattice 
artifacts can be found in Ref.~\cite{lss}.

\section{Results} 

\begin{figure}[h!]
\includegraphics[width=0.9\textwidth]{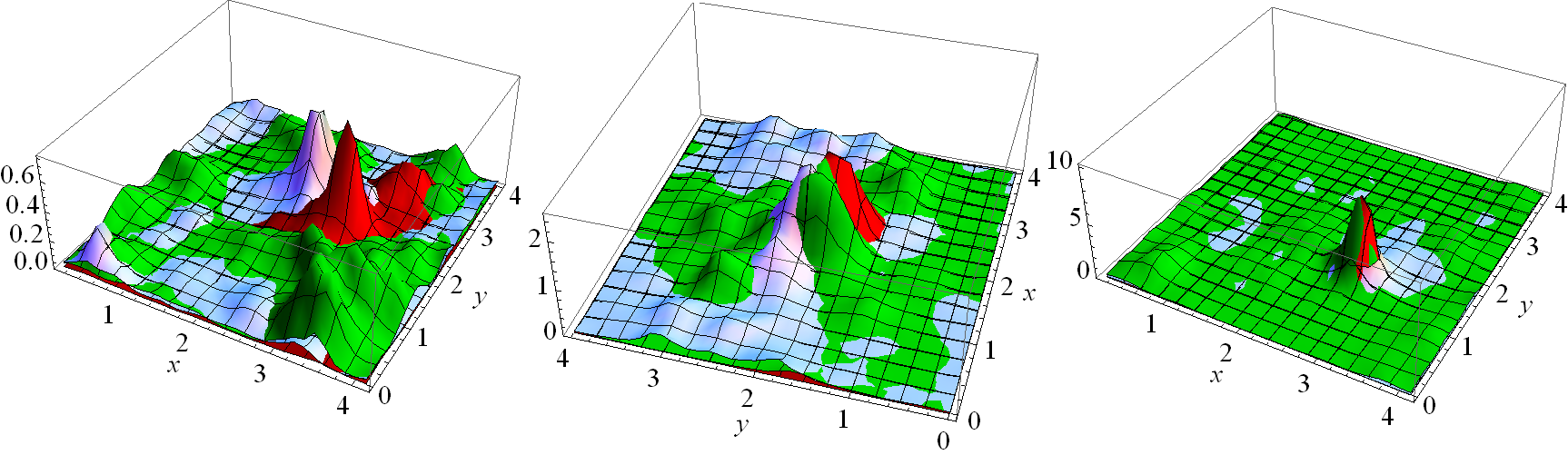}
 \caption{Wavefunction density $\rho (x,y)$ of the overlap fermion zero-mode summed over the temporal direction of three statistically independent QCD configurations at $T_c$ (left panel) and at $T= 1.08~T_c$ (middle and right panels). The three different colors represent 
the zero-modes at temporal periodicity phases $\phi = \pi$ (red), $\phi = \pi /3$ (blue), $\phi = -\pi /3$ (green) respectively. Peak heights have been normalized corresponding to the height for $\phi = \pi$. The $x$ and $y$ coordinates are in units of $1/T$.}
\label{fig:zeromodeprofile}
\end{figure}

As mentioned earlier, we focus on $|Q|=1$ configurations, with one overlap fermion zero-mode each, 
for 3 choices of the periodicity phase. 
The snap-shots of the eigenvalue density $\rho(x,y)$ along two spatial directions 
exemplified in Fig.~\ref{fig:zeromodeprofile} provide extreme cases of well separated and 
strongly overlapping topological objects. These zero modes display rich underlying topological 
structures of the QCD vacua at finite temperatures, both at $\sim T_c, 1.08~ T_c$.
The profiles for three different fermion periodicity phases  $\phi=\pi,-\pi/3$ and $\pi/3$ are shown in red, green and blue colors respectively. The left panel represents a case at $T\sim T_c$ when the observed topological objects are reasonably well separated in space, making their individual identification possible by changing the periodicity phase of the valence (overlap) fermion operator. The positions of the zero modes shift as the periodicity phases of fermions are changed. This provides us an evidence that the objects we observe are dyons.  
Next we study the vacuum profiles of two statistically independent QCD configurations at $1.08~T_c$, shown in the middle and the right panels of Fig.~\ref{fig:zeromodeprofile} respectively. In the middle panel, the fermion zero-modes can be seen to be localized at different spatial coordinates when the periodicity conditions are changed, like in the previous configuration near $T_c$. The configuration shown in the right panel shows a different limiting case where all three zero-modes are superimposed at the same location.

We compared the observed eigenmode density with the analytic expression for the dyon
zero-mode density $\rho(x)=-1/(4 \pi^2)\partial_x^2 f_x(\phi,\phi)$ along four-vector $x$ as calculated in Ref.~\cite{Chernodub:1999wg}, with the function $f_x$ defined by
\begin{equation}
\label{eq:ZeroMode}
 \left(D_\phi^2+r^2(\mathbf x,\phi)+\sum_{m=1}^3\delta_m(\phi)\right)f_x(\phi,\phi')=2 \pi \delta(\phi -\phi')~.
\end{equation}
Here $D_\phi=\frac{1}{i}\partial_\phi-\tau$ and the function $r^2(\mathbf x,\phi)=r_m^2(\mathbf x)$, if the fermion periodicity phase $\phi\in[\mu_m,\mu_{m+1}]$ i.e. lies between the two neighboring holonomy phases. The delta function chooses dyons with different holonomy
as evident from its definition, $\delta_m=\delta(\phi-\mu_m)\vert \mathbf {x_m-x_{m+1}}\vert$.
In fact the long-distance fall-off of the eigenvalue density $\rho(x)$ along any direction $x$ is controlled by the holonomy phases $\mu_m$ and the distance between dyons. When 
dyons are isolated and well separated from each other their long-distance fall-off is exponential in the difference between the periodicity phase $\phi$ and the nearest $\mu_m$. 

\begin{figure}[h!]
\begin{center}
\includegraphics[width=0.6\textwidth]{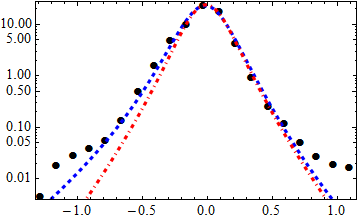}
\caption{The density profile of the quark zero-mode $\rho(x)$ at $\phi = \pi $ (black dots) for the configuration at $T=1.08~T_c$ discussed earlier in the right panel of Fig.~\ref{fig:zeromodeprofile} is compared to the analytic result with corresponding values for the Polyakov loop $P=0.4$ (blue dashed) and $P=1$ (red dot-dashed). The $x$-coordinate is scaled by $1/T$.
The peak shown in red (dot-dashed) has been scaled to fit the height of the lattice data, while blue (dashed) peak uses the found normalization. The position of other dyons for the blue profile are at $(x,y,z)\equiv(0.13/T,0.1/T,0.0)$ and $(0.1/T,-0.1/T,0.0)$ and for the red profile at $(0.14/T,0.0,0.0)$ and $(-0.14/T,0.0,0.0)$ respectively.}
\label{fig:ProfilefallOff}
\end{center}
\end{figure} 

In the case of well separated dyons, comparison with analytic formulae
is straightforward and successful. But even in a strongly overlapping case, as depicted 
in the right panel of Fig.~\ref{fig:zeromodeprofile}, we propose here a method which 
could be successfully used to distinguish between the two possibilities:  whether it 
represents an isolated caloron with trivial holonomy (all $\mu_{1,2,3}\rightarrow 0,\langle P\rangle=1$),  or closely superimposed dyons corresponding to non-trivial holonomy. To address this we show the eigendensity $\rho(x)$ along a spatial direction $x$ in Fig.~\ref{fig:ProfilefallOff} with the other 3 coordinates fixed at its maximum.  The lattice data are compared to the density calculated analytically for both of these options, for which the corresponding Polyakov loop values were chosen to be $\langle P\rangle=1$ (red curve) and $\langle P\rangle=0.4$ (blue curve) respectively. 

The zero-mode density is analytically calculated by placing the dyon associated with it at the origin. The height and shape of it are then controlled by the location of the two other dyons, and the value of the Polyakov loop. For $\langle P\rangle=1$, the two other dyon profiles are infinitely wide with zero heights. Therefore these only affect the size of the central dyon making it delocalized along the $\tau$-direction, as it should, since it reduces to the caloron solution. For $\langle P\rangle\neq 1$, the position of the other two dyons control not only the height and $\tau$-dependence, but also the direction along which the central dyon profile is tilted. As the two dyons approach the main dyon, its corresponding eigendensity becomes more localized along the $\tau$-direction with an increasing height. Moreover presence of a close dyon neighbor on any one of the sides of the main dyon causes the zero-mode solution to tilt along that direction with a larger slope, when we look at its one dimensional profile. If 
the zero-mode solution is not symmetric about its maxima, it thus gives a stronger indication that  it represents a dyon and not a caloron.

The eigendensity profile measured on the lattice, shown as points in Fig.~\ref{fig:ProfilefallOff} is not symmetric. For $\langle P\rangle=1$, the corresponding 
analytic solution is always symmetric about the maximum; while we could tune the height fairly accurately, it could not mimic the entire profile, specially the tails. For $\langle P\rangle=0.4$, the analytic profile of a dyon in presence of other localized dyons placed in non-symmetrical way, mentioned in the caption, could mimic the lattice profile quite well. We conclude that the long-distance fall-off of the density profile measured on the lattice is in better agreement with the blue rather than the red curve; so even the extreme case depicted in the right panel of Fig.~\ref{fig:zeromodeprofile} represents three overlapping dyons, rather than a caloron with trivial holonomy. 

\begin{figure}[h!]
\begin{center}
\includegraphics[width=0.65\textwidth]{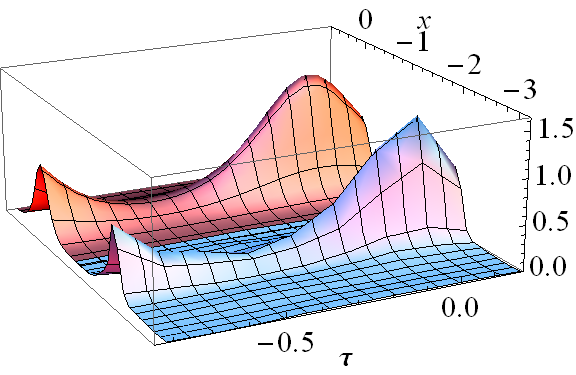}
\caption{Wavefunction density $\rho(x,\tau)$ at $\phi = \pi /3$ of the fermion zero-mode measured on the lattice in the $x-\tau$ plane (blue, right) compared to the analytic formula (red, left). The coordinates are scaled by $1/T$.}
\label{fig_2000_1pi3_xt}
\end{center}
\end{figure} 

The robustness of the semi-classical correspondence to the QCD vacuum is tested in 
another contrasting scenario at $1.08~T_c$ when the fermion zero modes are well 
separated. In this case, the lattice result for the zero-mode density 
$\rho(x,\tau)$ can provide a distinct indication for the distance between dyons.
The $\rho(x,\tau)$ for phase angle $\phi=\pi/3$ measured on the lattice is shown in blue 
in Fig.~\ref{fig_2000_1pi3_xt}.  Again modeling this analytically using Eq.~\ref{eq:ZeroMode} with $\langle P\rangle=0$ and placing the three dyons corresponding to $\phi=\pi/3,-\pi/3$ 
and $\pi$ respectively at coordinates $(0,0,0), (0.2/T,0,0)$ and $(-0.2/T,0,0)$, we obtained 
the profile shown in red in the same figure which mimics the lattice result quite well. 
While the shapes and the amount of overlap between the peaks varies across configurations 
studied here, we have not found any instance were these could not be explained by the 
semi-classical theory of dyons.

\subsection{The near-zero modes of the QCD Dirac operator}
The study of near-zero modes of the QCD Dirac operator can provide further insights on 
the interactions between topological constituents of QCD vacuum. If the density of 
topological objects is small or they are fairly dense but weakly interacting, in either case 
results in a very sparse near-zero eigenvalue density. 

\begin{figure}[h!]
\begin{center}
\includegraphics[width=0.95\textwidth]{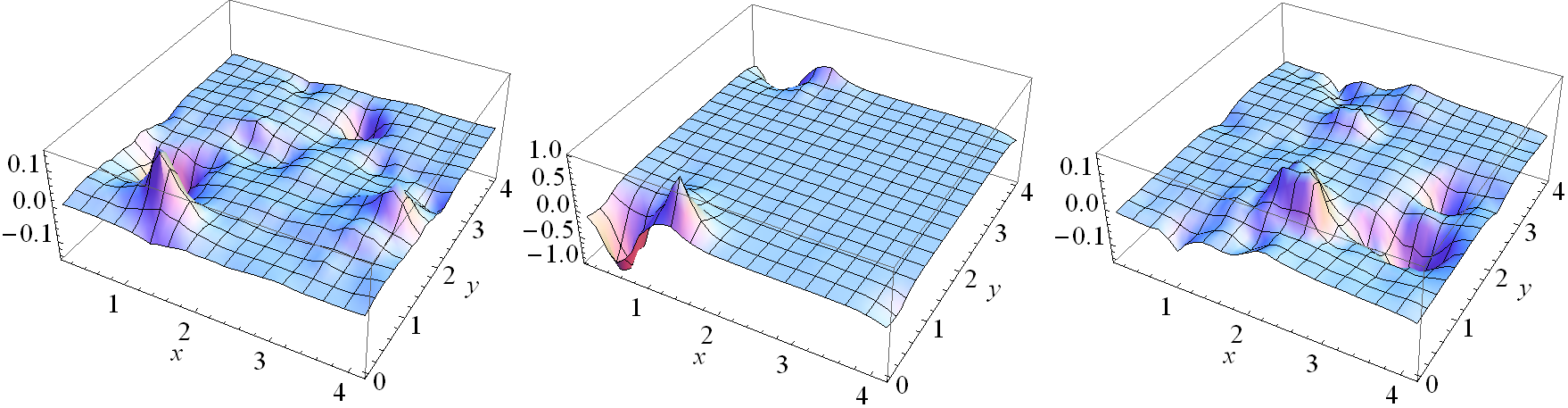}
\caption{Chiral density $\rho _5(x,y)$ profiles of the first near-zero mode for quark periodicity phase $\phi = \pi$ at $T_c$ (left panel) and for $\phi=\pi$ (middle) and $\phi= \pi/3$ (right panel) for $T=1.08~T_c$. These are the same QCD configurations whose zero-modes were shown earlier in left and middle panels of Fig. ~\ref{fig:zeromodeprofile}. The coordinates are scaled by $1/T$.}
\label{fig:NearZeroModes}
\end{center}
\end{figure}

We look into the first near-zero mode for a configuration at $1.08~T_c$ presented earlier in the middle panel of Fig.~\ref{fig:zeromodeprofile}. In order to extract maximum information out of the near-zero modes, their chiral densities are measured at two different periodicity phases $\phi=\pi$ and $\phi=\pi/3$ respectively, results of which are shown in middle and right panels of Fig. \ref{fig:NearZeroModes}. For $\phi=\pi$, we observe one pair of closely located dyon and an anti-dyon. Changing to $\phi=\pi/3$ a more richer picture emerges; the dominant dyon that contributed to the zero-mode has a prominent peak with positive chirality and atleast two distinct anti-dyon peaks are located near to it. These results have a very nice physical interpretation. Since the underlying gauge configurations were generated with anti-periodic Matsubara periodicity conditions for the fermions, the fermion determinant naturally suppressed those configurations which consisted of well separated $L$-dyons. In contrast, configurations with well separated $M$-dyons (corresponding to the phase $\phi=\pi/3$ ) or $M\bar M$ pairs are 
not suppressed. Moreover it is expected that at $T>T_c$ the $M$-dyons are lighter than the $L$-dyons and hence are more numerous. We next look at the near-zero mode profiles for the QCD configuration closer to $T_c$ for $\phi=\pi$ whose zero modes were observed earlier in the left panel of Fig.~\ref{fig:zeromodeprofile}. The density of $L\bar L$ dyons is believed to increase as one approaches $T_c$ leading to \emph{collectivization} of near-zero modes of the QCD Dirac operator. This is also what we observe in the left panel of Fig. \ref{fig:NearZeroModes}. 
Unlike calorons, dyons interact directly with the holonomy potential. Increasing density 
of dyons near $T_c$ is long suspected to provide a strong back-reaction to overcome the perturbative Gross-Pisarski-Yaffe potential~\cite{Gross:1980br} for the Polyakov loop and drive it towards its confining value. This first principles observation of \emph{collectivization} in realistic QCD configurations would eventually help towards an understanding of the mechanism of confinement.

\section{Summary}
Unlike various versions of ``cooling", the fermion eigenmode method we use
reveals the underlying topological objects in QCD configurations without any 
modifications. For the first time, we applied this to QCD configurations 
with physical quarks and with chiral symmetry realized to a good precision on the 
lattice.  Our choice of valence overlap fermions with an index theorem allowed us to 
study the properties of topological objects at temperatures $T_c$ and $1.08~T_c$, 
focusing on its zero modes. Changing the Matsubara periodicity phases of the
(overlap) fermions to $\phi=\pi,\pi/3,-\pi/3$, we identified all three types of 
dyons by observing their locations and their density profiles. Though 
use of different temporal periodicity phases of the fermion 
 operator to probe the presence of dyons is not new~\cite{Bornyakov:2015xao}, our 
 detection strategies are distinctly different.  In contrast to Ref.~\cite{Bornyakov:2015xao}, 
 where dyons are identified as local topological charge clusters, constructed out of a lowest 
 few eigenvalues and vectors of valence Dirac operator with different temporal boundary phases, 
 we actually monitor the global profiles of the  exact fermion zero modes and its shift in position  due to the change in the periodicity phases.  We have shown that the fermionic zero-modes method is perhaps the best to extract the topological content of QCD vacuum. If the chiral symmetry of the fermions are preserved to a very high precision on the lattice, it is possible to identify the topological objects using them even in the presence of strong quantum and thermal fluctuations near $T_c$. The other advantage of this method is that no gauge fixing is required to detect dyons which carry both color electric and magnetic charges.

We extensively studied different possible situations, ranging from widely separated dyons to 
fully overlapping ones. By quantitatively comparing their spacetime profiles with the available 
analytic formulae for well separated and, most importantly, partially overlapping dyons, we 
found in all cases a good agreement, with deviations not exceeding $\sim10$-$20\%$ level.
This is remarkable accuracy for a semi-classical theory, taking into account the 
fact that the typical action per dyon is $S\sim 3$-$4~\hbar$ and naively, relative 
fluctuations are expected to be $O(1/S)$. Our work thus provides a first evidence 
that the semi-classical description of dyons is possible already above $T_c$, in a 
deeply non-perturbative regime of QCD.

Furthermore we could extract rich qualitative information about the density and interactions between dyons from a detailed study of chiral density profiles of quark near-zero modes. At 
$T_c$, the density of $L$-dyons increases and the attractive interaction between $L\bar L$ 
pairs lead to \emph{collectivization} of near-zero modes. As the temperature is increased, the 
density of dyons decreases resulting in a sparse near-zero eigendensity of the QCD Dirac operator.
We are continuing this work towards a more quantitative measurement of the density and correlations 
between the dyons in a forthcoming publication~\cite{lss}.

\section{Acknowledgments}
This work was supported in part by the Office of Science, U.S. Department of Energy,   
under Contract Nos. DE-FG-88ER40388 (E.S) and DE-SC0012704 (R.N.L) and by the Department of 
Science and Technology, Govt. of India through a Ramanujan fellowship (S.S). We thank the 
HotQCD collaboration, formerly also consisting of members from the RBC-LLNL collaboration, 
for sharing the domain wall configurations with us. S.S. gratefully acknowledges the support 
from the International Centre for Theoretical Sciences (ICTS), Bengaluru during a 
visit for participating in the program ICTS/Prog-NUMSTRINGS2018/01 during the early 
stages of this work. The computations were performed in the GPU cluster at the Institute 
of Mathematical Sciences. We thank P. Mangalapandi for technical support. Our GPU code is 
in part based on some publicly available QUDA libraries~\cite{Clark:2009wm}.

\end{document}